\documentclass[12pt]{iopart}

\usepackage{iopams}
\usepackage{braket}
\usepackage{amsfonts}
\usepackage{color}
\usepackage[dvipsnames]{xcolor}
\usepackage{amssymb}
\usepackage{mathrsfs}
\usepackage{graphicx}
\usepackage{lmodern}
\usepackage{lineno}
\usepackage{hyperref}
\usepackage{soul}

\newcommand{\beq}{\begin{eqnarray}}
\newcommand{\eeq}{\end{eqnarray}}
\newcommand{\nn}{\nonumber}
\newcommand{\lanln}[1]{\texttt{arXiv:#1}}

\def\keywords#1{\vspace{10pt}
     \begin{indented}
     \item[]\rm Keywords: #1\par
     \end{indented}}

\begin{document}


\title{The Unruh effect for higher derivative field theory}
\author{Jasel Berra--Montiel$^{1}$, Jairo Mart\'inez--Montoya$^{2}$ and 
Alberto Molgado$^{1,3}$}

\address{$^{1}$ Facultad de Ciencias, Universidad Autonoma de San Luis 
Potosi \\
Av.~Salvador Nava S/N Zona Universitaria, San 
Luis Potosi, SLP, 78290 Mexico}
\address{$^{2}$ Unidad Academica de Fisica, Universidad Autonoma de Zacatecas, Calzada Solidaridad esq. Paseo La Bufa S/N, Zacatecas, Zac, 98060 Mexico}
\address{$^3$ Dual CP Institute of High Energy Physics, Mexico}

\eads{\mailto{\textcolor{blue}{jberra@fc.uaslp.mx}},\ 
\mailto{\textcolor{blue}{molgado@fc.uaslp.mx}}\ 
}

\begin{abstract}
We analyse the emergence of the Unruh effect within the context of a field Lagrangian
theory associated to the Pais-Uhlenbeck fourth order oscillator model.
To this end, we introduce a transformation that 
brings the Hamiltonian bounded from below and 
is consistent with $\mathcal{PT}$-symmetric
quantum mechanics.  We find that, 
as far as we consider different frequencies 
within the Pais-Uhlenbeck model, a particle
together with an antiparticle of 
different masses are created as may be traced back to 
the Bogoliubov transformation 
associated to the interaction between the Unruh-DeWitt detector and the higher derivative scalar field.  
On the contrary, whenever we consider the 
equal frequencies limit, no particle creation is detected as the pair particle/antiparticle
annihilate each other.    Further,  following Moschella and Schaeffer, we construct a Poincar\'e invariant two-point function for the Pais-Uhlenbeck model, which in turn allows us to 
perform the thermal analysis for any of the 
emanant particles.
\end{abstract}




\section{Introduction}

Lagrangians depending on higher order derivatives appear in 
many areas of physics, including string theory, brane theories, and remarkably gravitation-like models for which 
the higher order dependence is introduced by means of  curvature 
terms as, for example, in the Regge-Teitelboim model~for gravity \`a la string~\cite{rt},~\cite{molgado-prd}, and Dirac's membrane model~\cite{tucker},~\cite{molgado-cqg}.  One may even encounter this kind of higher order derivative Lagrangians in
economy and biology~\cite{udriste},~\cite{tu}. In particular, in the 
gravitational context, one may consider Weyl's conformal gravity which is a theory of gravity constructed from the Weyl tensor.  Among the peculiarities of conformal gravity 
we emphasize that it contains higher order derivative terms and also it is a renormalizable theory. Despite of the importance of higher 
order theories in physics, the study of this kind of systems was slighted as there were some evidence on the 
appearance of associated negative norm states at the quantum 
level, commonly known as \textit{ghosts}.   However, in the 
past few years, higher order theories were analysed 
from different perspectives, among them the $\mathcal{PT}$-symmetric Hamiltonian version, from which the ghosts may disappear under suitable conditions on the physical 
states~\cite{bender1},~\cite{bender2},~\cite{bender3}. Indeed, in $\mathcal{PT}$ quantum mechanics if the Hamiltonian is not Dirac Hermitian, but instead is symmetric under parity reflexion and time reversal transformations, a new dynamical inner product must be introduced. This inner product is constructed by means of isospectral similarity transformations on the $\mathcal{PT}$-symmetric Hamiltonian, resulting in a positive definite Hamiltonian, whose eigenstates have a positive inner product.
In this sense, $\mathcal{PT}$-symmetric Hamiltonian quantization was introduced for the Pais-Uhlenbeck fourth order oscillator, avoiding the emergence of ghosts~\cite{bender1}.  
Besides,  by considering a different perspective, in references~\cite{pavsic1},~\cite{pavsic2},~\cite{pavsic3} the viability of higher derivative theories is studied by focusing on the analysis of their instabilities even if the Hamiltonian results not bounded from below.

In this paper we present a field theoretical version of the Pais-Uhlenbeck oscillator in order to analyse the appearance of 
the Unruh effect, that is, the creation of particles as seen 
from an accelerated observer and corresponding to a thermal bath with temperature proportional to the acceleration.  Pais-Uhlenbeck oscillator has been proposed
as a toy model to study certain issues in conformal 
gravity.  Indeed, by considering a linear approximation around a flat spacetime 
background on the conformal action leads to a classical fourth order derivative 
theory, as discussed in~\cite{mannheim}, and also, symmetries of the Pais-Uhlenbeck model have been studied from the 
perspective of one of the subgroups of the conformal group~\cite{andre}. 
In consequence, 
our main aim is to address the emergence of the Unruh effect
associated to a higher order derivative theory. Our construction considers a 
transformation which brings the Pais-Uhlenbeck field model to a well-defined bounded from below Hamiltonian when 
decomposed into Fourier modes. This decomposition is in full
agreement with the principles developed within $\mathcal{PT}$-symmetric quantum mechanics~\cite{bender-pt}.
Further, we note that the Unruh effect only appears in the 
case that the frequencies that characterize 
the Pais-Uhlenbeck 
oscillator are unequal.  Within our setup, this can be easily 
seen as the coefficients appearing in the Bogoliubov 
transformation vanish in the equal frequency limit.  This 
confirms the intuitive picture as for the equal frequency 
limit the Pais-Uhlenbeck model is known to have continuous
eigenvalues, in opposition to the discrete eigenvalues 
obtained for unequal frequencies~\cite{bolonek},~\cite{pudeformation}.
As for the thermal behaviour, we adapt Moschella and 
Schaeffer construction~\cite{moschella1}, \cite{moschella2}, 
to the model of our interest.  Due to the symmetry of 
the Pais-Uhlenbeck decomposition into modes, we argue
that the two-point function describing the system behaves  
as a couple of subsystems, one for a particle and one for an antiparticle, for which the temperature is associated to 
a factor proportional to their different masses.


The rest of the article is organized as follows.  
In Section \ref{sec:PUfield} we start by describing the Lagrangian and Hamiltonian formalisms for the Pais-Uhlenbeck field model. In 
Section~\ref{sec:Unruh} we analyse extensively the Unruh effect adapted to our original setup. In opposition to the 
standard analysis of the Unruh effect, we
discuss the creation of a pair of particles of different masses.  Finally, in Section~\ref{sec:conclu} 
we include some concluding remarks.

\section{Pais-Uhlenbeck field model}
\label{sec:PUfield}

We consider an action described in a four dimensional Minkowski spacetime with 
signature $(+---)$ corresponding to the field version of the Pais-Uhlenbeck oscillator~\cite{jimenez} with a current term, representing the Unruh-DeWitt detector, 
\beq\label{PU action}
S&=&\int d^4x \left[\frac{1}{2}\phi(\Box + {m_1}^2)(\Box + {m_2}^2)\phi + J\phi \right] \nonumber\\
&=&\int d^4x \left[ \frac{1}{2}(\Box\phi)^2 - \frac{1}{2}({m_1}^2+{m_2}^2)\partial_{\mu}\phi\partial^{\mu}\phi + \frac{1}{2}{m_1}^2{m_2}^2{\phi}^2 + \epsilon Q \phi \right]\,,
\eeq
where the first term in the first line represents the Pais-Uhlenbeck oscillator while the second term stands for the Unruh-DeWitt detector, from the last line we identify $J=\epsilon Q$, $\epsilon$ being a free parameter associated to the detector efficiency and $Q=Q(t)$ being an arbitrary function described by a point-like object endowed with an internal structure that couples with a scalar field by means of a monopole moment, as described in~\cite{dewittqft},~\cite{takagi},~\cite{birrell}.  The function $\phi=\phi(t,\vec{x})$ stands for the 
Pais-Uhlenbeck scalar field, and the {\it masses} $m_1$
and $m_2$ are positive definite parameters characterizing 
the system. We emphasize that, in order to analyse the 
Unruh effect, we will consider that the current term $J$ 
is only active for a certain period of time, that is,
\beq
\label{eq:current}
J(t)& := &\left\{ \begin{array}{ll}
           0 & t < 0\\
           \epsilon Q(t) & 0< t< \tau\\
           0 & t>\tau
           \end{array}
           \right.\,.
\eeq 
Despite of the arbitrariness of the function $Q(t)$, we
will only focus our attention in the regions before and after it interacts
with the field $\phi(x)$.
In the regions without interaction $J(t)=0$, and by expanding the field $\phi(x)$ in spatial Fourier modes
\begin{eqnarray}
\phi(x)=\int\frac{d^3k}{(2\pi)^{3/2}}\phi_k e^{i \vec{k}\cdot\vec{x}}\,,
\end{eqnarray} 
the field equation of motion 
\beq
(\Box+m_1^2)(\Box+m_2^2)\phi = 0 
\eeq
reduces to
\begin{eqnarray}
0 & = & \int d^3k \frac{e^{i \vec{k}\cdot\vec{x}}}{(2\pi)^{3/2}} \left[ \frac{d^4\phi_k}{dt^4} + (2k^2+{m_1}^2+{m_2}^2)\frac{d^2\phi_k}{dt^2} \right.    \nn\\
& & 
\left. + (k^4+k^2({m_1}^2+{m_2}^2)+{m_1}^2{m_2}^2)\phi_k \right] \,,
\end{eqnarray}
and from this expression we obtain, by comparison with the equation of motion of the Pais-Uhlenbeck oscillator, the dispersion relations 
\begin{eqnarray}
{\omega_1}^2
=
k^2+{m_1}^2 \,,
\qquad 
{\omega_2}^2
=
k^2+{m_2}^2\,,
\end{eqnarray}
where $\omega_1$ and $\omega_2$ stand for the frequencies of the Pais-Uhlenbeck oscillator.    We note that each Fourier mode represents an independent 
Pais-Uhlenbeck fourth order oscillator.
From now on, and without loss of generality, we will consider 
$\omega_1 > \omega_2$.  As we will see, the equal frequency limit results inadequately defined when obtained from the 
general perspective $\omega_1\neq\omega_2$ as the quantum 
behaviour results completely 
different~\cite{bolonek},~\cite{pudeformation}. 
As shown in these references, in the equal frequency limit case 
the quantum Hamiltonian may be decomposed into a part 
proportional to an angular momentum term with discrete 
spectrum and a part proportional to a vector norm term
with continuous spectrum, in opposition to the case of our interest for which only a discrete spectrum is obtained. 
However, by taking the equal frequency limit in our formulation we will have some insights that the Unruh effect will not emerge in this case.  Further, as discussed 
in~\cite{pudeformation}, even though for
the classical equal frequency  limit 
the canonical transformation diverges, at the quantum level the canonical transformation results 
well-defined in this limit allowing to pass the wave functions from the discrete to the continuous case, thus 
supporting our claims about the Unruh effect in the following sections.
We will include appropriate comments 
whenever necessary in order to clarify this issue. 

Since the Pais-Uhlenbeck field theory possesses higher derivative terms, in order to get the Hamiltonian, we proceed using the standard Ostrogradski formalism. To implement this method, the phase space involves, in addition to the canonical coordinates $(\phi,\pi_{\phi})$, an extra pair of canonical variables associated to its higher order nature, namely $\psi:=\partial_0\phi$, with corresponding canonical momentum $\pi_{\psi}$. The momenta are defined explicitly as
\begin{eqnarray}
\pi_{\phi}&=&\frac{\partial\mathscr{L}}{\partial(\partial_0\phi)}-\partial_0\left(\frac{\partial\mathscr{L}}{\partial(\partial_0^2\phi)}\right) =-\left[({m_1}^2+{m_2}^2)\partial_0\phi + \partial_0\Box\phi\right] \nonumber\\
\pi_{\psi}&=&\frac{\partial\mathscr{L}}{\partial(\partial_0\psi)} = \Box\phi\,,
\end{eqnarray}
where $\mathscr{L}$ denotes the Lagrangian density appearing in the action~(\ref{PU action}), and $\partial_0$ denotes the partial time derivative.  
The canonical Hamiltonian, $H:=\int d^3x\, \mathcal{H}$, is thus obtained in a standard manner through the Legendre transformation, $\mathcal{H} = \partial_0\phi\pi_{\phi} + \partial_0\psi\pi_{\psi} - \mathscr{L}$.  Hence, the 
Hamiltonian density reads
\beq
 \mathcal{H} 
&=& 
\psi\pi_{\phi} + \frac{1}{2}{\pi_{\psi}}^2 + \frac{1}{2}(m_1^2+m_2^2){\psi}^2 - \frac{1}{2}m_1^2 m_2^2{\phi}^2 
- \left( \partial_{\mu}\nabla^2\phi \right)\partial^{\mu}\phi 
\nn \\
& &   
 - \frac{1}{2}(m_1^2+m_2^2)\nabla\phi \cdot \nabla\phi +
\partial_{\mu}\left( \nabla^2\phi\partial^{\mu}\phi \right)
- \epsilon Q\phi  \,,
\eeq
where the next to last term  corresponds to a boundary term.  If we expand the field into spatial Fourier modes we get independent Pais-Uhlenbeck oscillators interacting 
with the Unruh-DeWitt detector for each mode 
\begin{eqnarray}\label{hamiltonian plus current}
\hspace{-10ex}
H = \int \frac{d^3k}{(2\pi)^{3/2}} \left[ \psi_{\vec{k}}\, \pi_{\phi_{\vec{k}}} + \frac{1}{2}{\pi_{\psi_{\vec{k}}}}^2 + \frac{1}{2}({\omega_1}^2+{\omega_2}^2){\psi_{\vec{k}}}^2 - \frac{1}{2}{\omega_1}^2{\omega_2}^2{\phi_{\vec{k}}}^2 - \epsilon Q\phi_{\vec{k}} \right]\,,
\end{eqnarray}
as expected.

In the Fock space representation, we introduce two different pairs of creation and annihilation operators for each mode
\begin{eqnarray}
\label{eq:transformation}
\phi_{\vec{k}} &:=& \frac{1}{\Delta} \left[ \frac{1}{\sqrt{2\omega_2}}(b_{\vec{k}}+{b_{\vec{k}}}^{\dagger})-i\frac{1}{\sqrt{2\omega_1}}({a_{\vec{k}}}^{\dagger}-a_{\vec{k}}) \right] \nonumber \\
\pi_{\phi_{\vec{k}}} &:=& \frac{\omega_1\omega_2}{\Delta} \left[ i\frac{\omega_1}{\sqrt{2\omega_2}}(b_{\vec{k}}-{b_{\vec{k}}}^{\dagger})-\frac{\omega_2}{\sqrt{2\omega_1}}({a_{\vec{k}}}^{\dagger}+a_{\vec{k}}) \right] \nonumber \\
\psi_{\vec{k}} &:=& \frac{1}{\Delta} \left[ \sqrt{\frac{\omega_1}{2}}({a_{\vec{k}}}^{\dagger}+a_{\vec{k}})-i\sqrt{\frac{\omega_2}{2}}(b_{\vec{k}}-{b_{\vec{k}}}^{\dagger}) \right] \nonumber \\
\pi_{\psi_{\vec{k}}} &:=& \frac{1}{\Delta} \left[ i\omega_1\sqrt{\frac{\omega_1}{2}}({a_{\vec{k}}}^{\dagger}-a_{\vec{k}})-\omega_2\sqrt{\frac{\omega_2}{2}}(b_{\vec{k}}+{b_{\vec{k}}}^{\dagger}) \right]
\end{eqnarray}
where $\Delta:=\sqrt{\omega_1^2-\omega_2^2}$. These operators satisfy the equal time commutation relations 
\beq
\label{eq:basiccomm}
[ a_{\vec{k}},a_{\vec{k}'}^{\dagger} ] 
= 
\delta(\vec{k}-\vec{k}')
=
- [ b_{\vec{k}},b_{\vec{k}'}^\dagger ] \,, 
\eeq 
and, otherwise they vanish, 
The choice of the sign in front of the commutation relation between the $b_{\vec{k}}$ and $b^{\dagger}_{\vec{k}}$ operators is
necessary in order to guarantee that the energy spectrum of the Hamiltonian is real and  bounded from below~\cite{bender}.

In terms of the creation and annihilation operators~(\ref{eq:transformation}), the Hamiltonian for the Pais-Uhlenbeck model is given by
\begin{eqnarray}\label{hamiltonian with creation ops}
H &=& 
\int \frac{d^3k}{(2\pi)^{3/2}} \left[ \omega_1\left({a_{\vec{k}}}^{\dagger}a_{\vec{k}} + \frac{1}{2}\right) - \omega_2\left({b_{\vec{k}}}^{\dagger}b_{\vec{k}} - \frac{1}{2}\right) 
\right.  \nn\\
& & 
+ \left. 
i \epsilon \frac{Q}{\Delta}\frac{({a_{\vec{k}}}^{\dagger} - a_{\vec{k}})}{\sqrt{2\omega_1}} - \epsilon \frac{Q}{\Delta}\frac{(b_{\vec{k}} + {b_{\vec{k}}}^{\dagger})}{\sqrt{2\omega_2}} \right]\,,  
\end{eqnarray}
The first two terms in this Hamiltonian correspond to the diagonal operator representation of an infinite sum of Pais-Uhlenbeck oscillators, while the 
terms in the second line  comprise the different way in 
which the Unruh-DeWitt detector interacts with the 
introduced creation and annihilation operators. In case $\epsilon=0$, by considering the basic commutators~(\ref{eq:basiccomm}), and the property that $a_{\vec{k}}$ and $b_{\vec{k}}$ annihilate the 0-particle state $\left|\Omega\right>$,
\begin{equation}
a_{\vec{k}}\left|\Omega\right>=0 \,, 
\qquad
b_{\vec{k}}\left|\Omega\right>=0 \,,
\end{equation}
we can observe that the state $\left|\Omega\right>$ corresponds to the ground state with energy $\frac{1}{2}(\omega_{1}+\omega_{2})$ for each mode.
As noted in~\cite{bender}, although the Hamiltonian (\ref{hamiltonian with creation ops}) is not manifestly Dirac Hermitian, it results to be invariant under $\mathcal{PT}$ transformations. This means, that the ghost states can be reinterpreted as positive quantum states with respect to a suitable $\mathcal{PT}$ inner product. This product is constructed by introducing a dynamical reflection symmetry operator $\mathcal{C}$, which commutes with the Hamiltonian and the $\mathcal{PT}$ operator. Then, under this new inner product the norm of the states prove to be strictly positive. Similarly, in order to analyse the thermal behaviour of the Pais-Uhlenbeck particles, we will show that a positive inner product defined on the space of solutions must be introduced with the aim of avoiding the emergence of ghosts.
\\        
Finally, we note that the term 
$\Delta=\sqrt{\omega_1^2-\omega_2^2}$ appearing in~(\ref{eq:transformation}) becomes singular in the 
equal frequencies limit, this limit corresponds to the case of equal masses for a given mode. As stated 
in~\cite{bolonek},~\cite{pudeformation}, in the equal frequencies case, the 
behaviour of the Pais-Uhlenbeck system results completely 
different as  the spectrum of the Hamiltonian is composed of a discrete spectrum part coming from an angular momentum term, and of a continuous part originated from a different term given by the squared norm of the position variables.

\section{The Unruh effect for the Pais-Uhlenbeck field}
\label{sec:Unruh}

In this Section we will explore the Unruh effect  construction 
for the field theory developed so far, demonstrating the 
way in which particles are created for such model, and we also explore the thermal behaviour of the system  by 
introducing an appropriate two-point function.

\subsection{Creation of particles and Unruh-DeWitt detector}

Following Heisenberg prescription, $i(df/dt)=[f,H]$, ($\hbar=1$), we may determine how the system evolves in time. In this way, using the commutation relations~(\ref{eq:basiccomm}), the evolution of the operators $a_{\vec{k}}$ and $b_{\vec{k}}$ is given by 
\begin{eqnarray}
\label{eq:evolution}
a_{\vec{k}}(t)&=&a_{\vec{k}}^{(in)}e^{-i\omega_1 t}+\frac{\epsilon}{\Delta\sqrt{2\omega_1}}\int_0^t e^{i\omega_1(t'-t)}Q(t')dt' \nonumber\\
b_{\vec{k}}(t)&=&b_{\vec{k}}^{(in)}e^{-i\omega_2 t}-\frac{i\epsilon}{\Delta\sqrt{2\omega_2}}\int_0^t e^{i\omega_2(t'-t)}Q(t')dt' \,,
\end{eqnarray}
where $a_{\vec{k}}^{(in)}$ and $b_{\vec{k}}^{(in)}$ are constants determined by the initial conditions at $t=0$.
Assuming that the current 
$J(t)=\epsilon Q(t)$ only acts in the interval $[0,\tau]$ 
(see definition~(\ref{eq:current})), 
we define the interaction terms 
\begin{eqnarray}
\label{eq:ja-jb}
J_a 
:=
\frac{\epsilon}{\Delta\sqrt{2\omega_1}}\int_0^\tau e^{i\omega_1 t'}Q(t')dt' \,,
\quad
J_b
:=
-\frac{i\epsilon}{\Delta\sqrt{2\omega_2}}\int_0^\tau e^{i\omega_2 t'}Q(t')dt'\,,
\end{eqnarray}
where the integrals are considered in the interval $[0,\tau]$
as only in this interval the current interacts with the Pais-Uhlenbeck field.  
Note that the non-symmetric manner in which $J_a$ and $J_b$ appear  in the expressions above results from the higher order structure of the model.
Next, we separate our operators $a_{\vec{k}}(t)$ and $b_{\vec{k}}(t)$ into two parts each, one before and one after the current interaction. Thus, we have
\begin{eqnarray}
a_{\vec{k}}(t)
=
\left\{ \begin{array}{ll}
a_{\vec{k}}^{(in)}e^{-i\omega_1 t}, & t < 0\\
a_{\vec{k}}^{(out)}e^{-i\omega_1 t}, & t>\tau
\end{array}
\right. \,,
\qquad
b_{\vec{k}}(t)
=
\left\{ \begin{array}{ll}
b_{\vec{k}}^{(in)}e^{-i\omega_2 t}, & t < 0\\
b_{\vec{k}}^{(out)}e^{-i\omega_2 t}, & t>\tau
\end{array}
\right. \,,
\end{eqnarray}
where the operators $a_{\vec{k}}^{(out)}$ and $b_{\vec{k}}^{(out)}$ are straightforwardly obtained from~(\ref{eq:evolution}) and~(\ref{eq:ja-jb}), and explicitly may be written as $a_{\vec{k}}^{(out)}:=a_{\vec{k}}^{(in)} + J_a$, and $b_{\vec{k}}^{(out)}:=b_{\vec{k}}^{(in)} + J_b$, respectively.

Before the interaction occurs we assume, for any value of $\vec{k}$ (thus, from now on we will omit the $\vec{k}$ index), that there exist
no particle states, ${\left|\Omega_{0,b}\right>}_{in}$ and
${\left|\Omega_{a,0}\right>}_{in}$, such that 
\begin{eqnarray}
\label{eq:invacuum}
a^{(in)}{\left|\Omega_{0,b}\right>}_{in}=0 \,,
\hspace{7ex}
b^{(in)}{\left|\Omega_{a,0}\right>}_{in}=0 \,,
\end{eqnarray}
where states $\left|\Omega_{0,b}\right>_{in}$ and $\left|\Omega_{a,0}\right>_{in}$ represent the absence of particles of type $a$ and $b$, respectively, while the state $\left|\Omega_{0,0}\right>_{in}$ stands for the Minkowski vacuum state  in the $t<0$ region. Once we have defined our vacuum, we can construct from the no particle state all the other states by repeated use of creation operators ${a^{(in)}}^{\dagger}$ and ${b^{(in)}}^{\dagger}$
\begin{eqnarray}
\label{eq:creation}
{\left|\Omega_{n,b}\right>}_{in}
=
\frac{1}{\sqrt{n!}}{({a^{(in)}}^{\dagger})}^n{\left|\Omega_{0,b}\right>}_{in} \,,
\qquad
{\left|\Omega_{a,m}\right>}_{in}
=
\frac{1}{\sqrt{m!}}{({b^{(in)}}^{\dagger})}^m{\left|\Omega_{a,0}\right>}_{in} \,.
\end{eqnarray}  
Analogously, after the interaction took place we also assume the existence of no particle states, $\left|\Omega_{0,b}\right>_{out}$ and $\left|\Omega_{a,0}\right>_{out}$, such that
\begin{eqnarray}
a^{(out)}{\left|\Omega_{0,b}\right>}_{out}=0 \,,
\qquad
b^{(out)}{\left|\Omega_{a,0}\right>}_{out}=0  \,,
\end{eqnarray}
where, similarly, states $\left|\Omega_{0,b}\right>_{out}$ and $\left|\Omega_{a,0}\right>_{out}$ represent the absence of particles of type $a$ and $b$, respectively, while the state $\left|\Omega_{0,0}\right>_{out}$ stands for the Minkowski vacuum state  in the $\tau<t$ region.  Of course, 
analogous construction to~(\ref{eq:creation}) occur for any state in the $\tau<t$ region.


As mentioned before we have two sets of states, 
${\left|\Omega_{a,b}\right>}_{in}$ corresponding to the 
region $t<0$, and ${\left|\Omega_{a,b}\right>}_{out}$ corresponding to the region $\tau<t$,
respectively.  As both sets are described in Minkowski spacetime, the idea is to compare both sets and see how they are related. This relation is described by the Bogoliubov transformation
\begin{eqnarray}
{\left|\Omega_{0,0}\right>}_{in}=\sum_{n,m}\Lambda_{n,m}{\left|\Omega_{n,m}\right>}_{out}
\end{eqnarray} 
where $\Lambda_{n,m}\in\mathbb{C}$. For the case of our interest, the explicit form of the Bogoliubov transformation reads 
\begin{eqnarray}
\label{eq:lambda}
\Lambda _{n,m}=\frac{1}{\sqrt{n!m!}}e^{-\frac{1}{2}\left( |J_a|^2+|J_b|^2 \right)} J_a^n J_b^m \,.
\end{eqnarray}
where $J_a$ and $J_b$ were defined in~(\ref{eq:ja-jb}).  
Indeed, by considering this Bogoliubov transformation one may easily verify relations~(\ref{eq:invacuum}) by taking into account the 
expressions $a^{(out)}=a^{(in)}+J_a$ and $b^{(out)}=b^{(in)}+J_b$ 
described above.

As time passes, one may wander if the state ${\left|\Omega_{0,0}\right>}_{in}$ may evolve until it reaches the state ${\left|\Omega_{a,b}\right>}_{out}$.  As in ordinary quantum mechanics, the probability of this to happen is
\begin{eqnarray}
\label{eq:rho}
\rho^2={\left|_{out}\!\left<\Omega_{a,b}|\Omega_{0,0}\right>_{in}\right|}^2={\left|_{out}\!\left<\Omega_{a,b}\right|\sum_{n,m}\Lambda_{n,m}\left|\Omega_{a,b}\right>_{out}\right|}^2={|\Lambda_{n,m}|}^2\,.
\end{eqnarray}
This result tells us that the probability that the vacuum state in the region $t<0$ evolves into a state with particles in the region $\tau<t$ may be different from zero.  Thus, we conclude that in the process  of interaction at the region $0<t<\tau$ particles were created. 
This particle creation is due to the interaction term with the Pais-Uhlenbeck field by means of the Unruh-DeWitt detector.  In order to associate this particle creation to the Unruh effect we have to calculate the transition probability of the Unruh-DeWitt detector from a ground state to an excited 
one~\cite{birrell}. By choosing an hyperbolic trajectory in Rindler coordinates, one may see that the transition probability is different from zero as far as we consider a massless 
Klein-Gordon field.  However, in the massive case this procedure results non-analytic~\cite{takagi}.
An alternative approach, developed in \cite{moschella1},~\cite{moschella2}, takes advantage of  the issue that we are dealing with systems possessing an  infinite number of degrees of freedom. Indeed, in this case one may construct uncountably unitarily inequivalent Hilbert space representations of the canonical commutation relations. Thus the choice of a specific representation depends on appropriate physical considerations such as renormalizability, thermal equilibrium, positivity conditions, etc. The thermal 
behaviour developed in next subsection will follow this latter approach by implementing a Poincar\'e invariant 
and positive definite representation necessary in order 
to obtain analytic expressions for the massive case which is an essential issue in our higher order derivative model.

Also note that the particle creation may be related to the discrete energy spectrum of the Hamiltonian as changes in the energy levels 
are interpreted as emission or absorption of  
particles.  On the contrary, for the equal frequencies 
limit, $\Delta\to 0$, the probability density $\rho$ goes to zero, 
as in this case the coefficients appearing in the Bogoliubov transformation vanish, that is, $\Lambda_{n,m} \to 0$. 
As a consequence, one may 
interpret this as the absence of the Unruh effect in 
the equal frequency limit.  Although, as proved 
in~\cite{pudeformation}, it is possible to construct 
a well defined equal frequency limit for the quantum 
canonical transformations, hence strengthening our claim above on the Unruh effect, a
more carefully analysis must be carry out from 
the very beginning
considering the equal frequency Pais-Uhlenbeck 
Hamiltonian for this case.

\subsection{Thermal radiation}

For the Pais-Uhlenbeck field model~(\ref{PU action}) we start by solving the field equation 
\begin{eqnarray}\label{pu eq}
(\Box+m_1^2)(\Box+m_2^2)\phi = 0 \,.
\end{eqnarray}
 which, by symmetry, may be obtained  by considering 
the linear complex combination $\phi(x)=\alpha u(x) + \beta v(x)$, where $\alpha,\beta$ are constants and $u(x)$  and $v(x)$ are independent solutions to Klein-Gordon equations with masses $m_1$ and $m_2$, respectively,~\cite{zwillinger}
\begin{eqnarray}
\label{eq:KGs}
(\Box+m_1^2)u(x)=0 \,,
\hspace{7ex}
(\Box+m_2^2)v(x)=0 \,.
\end{eqnarray}
From Pais-Uhlenbeck action~(\ref{PU action}), and by means of the Noether theorem, we define an inner product on the complex solutions to our field equations as
\begin{eqnarray}
\label{eq:innerp}
(\phi,\psi) &:=& i\int d\Sigma^{\mu} \left[ -(m_1^2+m_2^2)(\phi^*\partial_{\mu}\psi-\psi\partial_{\mu}\phi^*)+(\partial_{\mu}\phi^*\Box\psi-\partial_{\mu}\psi\Box\phi^*) \right. \nn\\
& &\left. -(\phi^*\partial_{\mu}\Box\psi-\psi\partial_{\mu}\Box\phi^*)\right] \,,
\end{eqnarray} 
where $d\Sigma^{\mu}=n^{\mu}d\Sigma$, with $n^{\mu}$ an unitary vector orthogonal to the spatial Cauchy hypersurface $\Sigma$, and $d\Sigma$ represents the volume element of $\Sigma$. Under this inner product the independent solutions $u(x)$ and $v(x)$ result orthogonal, that is, $(u,v)=0$.
In this manner, we can express the Pais-Uhlenbeck field operator as a formal expansion in terms of a family of complex independent classical solutions $\{u_{i}(x)\}$, and $\{v_{i}(x)\}$, in the following way
\begin{eqnarray}
\phi(x)=\sum_{i=0}^{\infty}\left[u_i(x)a_i + u_i^*(x)a_i^{\dagger} + v_i(x)b_i + v_i^*(x)b_i^{\dagger}\right]\,,
\end{eqnarray}
where $a$ and $b$ are the annihilation operators and $a^{\dagger}$ and $b^{\dagger}$ are the creation operators, respectively, and they follow the commutation relations~(\ref{eq:basiccomm}).

As discussed before, the 
ambiguity in the representations allows us to look for a general Poincar\'e invariant extension of the standard two-point function $W(x,x')=\left<\Omega\right|\phi(x)\phi(x')\left|\Omega\right>$ which, due to the symmetry of the Pais-Uhlenbeck model, is 
explicitly decomposed into modes as
\beq
\label{eq:twopointW}
W(x,x')=W_u(x,x')-W_v(x,x')  \,.
\eeq
The separation of the two-point 
function $W(x,x')$ into 
two-point  
functions for each mode resulted as a consequence of the orthogonality of the $u$ and $v$ modes within the inner 
product~(\ref{eq:innerp}).
In order to obtain the exact expression for the normalized $u$ and $v$ modes we need to solve the pair of 
Klein-Gordon equations~(\ref{eq:KGs}). We will solve these equations in Rindler spacetime in order to describe the 
Unruh effect. The coordinates for this spacetime are explicitly given 
by the mapping $(t,x,y,z)\mapsto (\rho\sinh\eta,\rho\cosh\eta,y,z)$.
Thus, focusing on the $\rho$ and $\eta$ components of the coordinates, 
the Klein-Gordon equations~(\ref{eq:KGs}) read
\begin{eqnarray}
\left( \frac{1}{\rho^2}\frac{\partial^2}{\partial \eta^2} - \frac{1}{\rho}\frac{\partial}{\partial \rho}\left(\rho\frac{\partial}{\partial \rho}\right) + m_1^2 \right)u 
& = &
0 \,,  \nn\\
\left( \frac{1}{\rho^2}\frac{\partial^2}{\partial \eta^2} - \frac{1}{\rho}\frac{\partial}{\partial \rho}\left(\rho\frac{\partial}{\partial \rho}\right) + m_2^2 \right)v 
& = &  
0 \,,
\end{eqnarray}
from which we obtain the wave-like solutions
\beq
\label{eq:uv-prod-omegas}
u(\eta,\rho) 
= 
e^{-i\omega_1\eta}K_{i\omega_1}(m_1\rho)  \,,
\qquad
v(\eta,\rho) 
=
e^{i\omega_2\eta}K_{i\omega_2}(m_2\rho) \,,
\eeq
respectively.  Note that the difference in signs in the exponential of the two modes is necessary to guarantee  
the positiveness of the inner product defined in~(\ref{eq:innerp}). Furthermore, this positivity condition can be traced back to the non-Hermiticity property of the Pais-Uhlenbeck Hamiltonian as, for these kind of systems, in order to ensure a positive inner product and normalized wave functions $\mathcal{PT}$-symmetric eigenstates must be introduced~\cite{bender}. Similarly, in the present case, these states are given by the solutions $v(\eta,\rho)$, which result to be positive definite under the inner product derived from the conserved current (\ref{eq:innerp})  
Note that in~(\ref{eq:uv-prod-omegas}), the functions $K_{i\omega_n}(m_n\rho)$, for $n=1,2$,
stand for the Macdonald functions~\cite{atlas} that are solutions to the 
modified Bessel equation of the second kind with imaginary index
\begin{eqnarray}
\label{eq:mac}
\hspace{-10ex}
\rho^2\frac{d^2 K_{i\omega_n}(m_n\rho)}{d \rho^2} + \rho\frac{d K_{i\omega_n}(m_n\rho)}{d \rho} - \left(m_n^2\rho^2-\omega_n^2\right)K_{i\omega_n}(m_n\rho)=0\,.
\end{eqnarray}
At this point, it is important to mention 
that we choose the solution 
$K_{i\omega_n}(m_n\rho)$ to~(\ref{eq:mac}) due to its 
asymptotic behaviour $K_{i\omega_n}(m_n\rho)\approx (\pi/2m_n\rho)^{1/2}e^{-m_n\rho}(1+O((m_n\rho)^{-1}))$ which tends to zero as the argument tends to infinity.  This issue is relevant in order to analyse the Unruh 
effect as we assume an adiabatic interaction for an static spacetime~\cite{birrell}.

We also note that, by considering the inner product~(\ref{eq:innerp}), the normalization of the $u$ and $v$ modes is given by
\begin{eqnarray}
\label{eq:innerp-uv}
(u_{\omega_1}(\eta,\rho),u_{\omega_1'}(\eta,\rho)) &=& 
\frac{\Delta^2\pi^2}{\sinh(\pi\omega_1)}\delta(\omega_1-\omega_1') \,,\nn\\
(v_{\omega_2}(\eta,\rho),v_{\omega_2'}(\eta,\rho)) &=& 
\frac{\Delta^2\pi^2}{\sinh(\pi\omega_2)}\delta(\omega_2-\omega_2') \,.
\end{eqnarray}
Again, we may note that at the equal 
frequencies limit $\Delta\to 0$ this product
automatically vanishes.  We interpret in a 
simple manner as in this limit the Pais-Uhlenbeck model is invariant under a different 
symmetry, and thus, the inner product for the equal frequency limit 
may be obtained from an appropriate 
Noether current.  In view of the inner products~(\ref{eq:innerp-uv}) we finally propose the normalized modes
\begin{eqnarray}\label{eq:modos uv}
u_{\omega_1}(\eta,\rho) &=& \frac{\sqrt{\sinh(\pi\omega_1)}}{\Delta\pi}e^{-i\omega_1\eta}K_{i\omega_1}(m_1\rho) 
\,,\nn \\
%
v_{\omega_2}(\eta,\rho) &=& \frac{\sqrt{\sinh(\pi\omega_2)}}{\Delta\pi} e^{i\omega_2\eta}K_{i\omega_2}(m_2\rho) \,.
\end{eqnarray}
The different signs of the exponential functions  may be interpreted  as particles of mass $m_1$ and  
antiparticles of mass $m_2$, respectively.
This interpretation is consequent with the equal frequency limit discussed above, as within this case
no particles are observed.  From this point of view, we may simply infer that in 
the equal frequency limit particles and antiparticles annihilate each other.

The generalized Poincar\'e invariant two-point functions in terms of the normalized modes are given by (see~\cite{moschella1},~\cite{moschella2} for details)
\begin{eqnarray}
\label{funcion modos uv}
\hspace{-10ex}
W_u(x,x') 
& = & 
\frac{1}{\pi^2\Delta}\int_0^{\infty} \sinh(\pi\omega_1)\left[ \frac{e^{-i\omega_1(\eta-\eta')}}{1-e^{-2\pi\omega_1}} -\frac{e^{i\omega_1(\eta-\eta')}}{1-e^{2\pi\omega_1}} \right] K_{i\omega_1}(m_1\rho)K_{i\omega_1}(m_1\rho')d\omega_1 \,, \nn \\
\hspace{-10ex}
W_v(x,x') 
& = & \frac{-1}{\pi^2\Delta}\int_0^{\infty}\sinh(\pi\omega_2)\left[  \frac{e^{-i\omega_2(\eta-\eta')}}{1-e^{-2\pi\omega_2}} -
\frac{e^{i\omega_2(\eta-\eta')}}{1-e^{2\pi\omega_2}} \right] 
K_{i\omega_2}(m_2\rho)K_{i\omega_2}(m_2\rho')d\omega_2 \,, \nn\\
& & 
\end{eqnarray}
The terms in the square brackets may be written 
as hyperbolic cosines, respectively, and thus, by considering the 
integral identity~\cite{bateman}
\begin{eqnarray}
\frac{2}{\pi}\int_0^{\infty}K_{iy}(a)K_{iy}(b)\cosh\left[ y(\pi-\phi) \right]dy = K_0\left[ \sqrt{a^2+b^2-2ab\cos(\phi)} \right] \,,
\end{eqnarray} 
we finally obtain the expressions
\begin{eqnarray}
\label{two  point function poincare}
W_u(x,x') 
& = &
\frac{1}{2\pi\Delta^2}K_0\left( m_1||x-x'|| \right) \nn \\
W_v(x,x') 
& = & 
-\frac{1}{2\pi\Delta^2}K_0\left( m_2 ||x-x'|| \right) \,,
\end{eqnarray}
where the terms $||\cdot||$ in the argument of the Macdonald $K_0$ functions 
stand for the Minkowski norm.  In this way, we have obtained the complete generalized Poincar\'e invariant two-point function~(\ref{eq:twopointW}) associated to the 
Pais-Uhlenbeck model
\begin{eqnarray}
\label{funcion a dos puntos completa}
\hspace{-7ex}
W(x,x')= \frac{K_0\left( m_1||x-x'|| \right) + K_0\left( m_2||x-x'|| \right)}{2\pi \Delta^2} \,.
\end{eqnarray}

Further, from~(\ref{funcion modos uv}) we note that any of the functions $W_u(x,x')$ and $W_v(x,x')$ may be regarded as the inverse Fourier transform of the functions $F(\omega_n),\ n=1,2$, identified as
\begin{eqnarray}
\label{eq:inverseFourier}
F(\omega_n)=\frac{1}{\pi^2\Delta^2}\, \frac{K_{i\omega_n}(m_n\rho)
K_{i\omega_n}(m_n\rho')\sinh(\pi\omega_n)}{e^{2\pi\omega_n}-1} \,.
\end{eqnarray} 
The function $F(\omega_n)$ thus represents the Fourier transform of the generalized two-point function 
$W(x,x')$. From the expression 
~(\ref{eq:inverseFourier}), we may see that the Planck factor $(e^{2\pi\omega_n}-1)^{-1}$ emerges naturally, thus concluding that 
each of our 
particles follows a Bose-Einstein distribution with  temperature related to their respective frequencies $\omega_n$. \\


\section{Conclusions}
\label{sec:conclu}

In this article, we analysed the Unruh effect that emerges from a  field theoretical version of the Pais-Uhlenbeck fourth order oscillator. Even though, higher derivative theories are commonly associated to the presence of ghost states and energy spectra that are not bounded by below resulting in the loss of unitarity, the Pais-Uhlenbeck oscillator do not suffer from these kind of disadvantages. This last result 
may be seen as a consequence of the discrete $\mathcal{PT}$-symmetry invariance of the Hamiltonian, obtained by introducing an appropriate dynamical inner product in order to preserve positive norm states, instead of using the standard Dirac inner product. Within this setup, we derived the Bogoliubov transformations associated to the interaction between the Unruh-DeWitt detector and the higher derivative scalar field. Afterwards, in order to describe the thermal behaviour present in the Unruh effect, our strategy was to obtain the most general Poincar\'e invariant two point-function. Due to the symmetries associated to the Pais-Uhlenbeck Hamiltonian, the quantum theory possesses a positive definite inner product and, therefore, unitarity is guaranteed. We also showed that, 
within the unequal frequencies case and as a consequence of the Bogoliubov transformation for the normalized modes,   particles of mass $m_1$ together with antiparticles of mass $m_{2}$ were created in the process. This interpretation results physically consequent with the equal frequency limit, as within this case the energy spectrum is now continuous, and no particles are observed at all. Finally, by using the generalized Poincar\'e invariant two point function, we conclude that each kind of the particles
created follow a Bose-Einstein distribution with temperature related to their respective frequencies, which in turn are associated to the value of their respective masses.
Further studies are necessary to guarantee that our 
affirmations on the absence of the Unruh effect in the 
equal frequency limit may be valid as we may consider a different Hamiltonian from scratch within this limit.  This construction effectively will change most of the structures involved as, for example, the inner product 
definitely corresponds to a different Noether current, thus affecting the two-point functions required for the
description of the thermal behaviour of the system.  
Also, a more general analysis must be carried on 
in order to 
discern on the generality of the results obtained here for a generic higher derivative model. This will be done elsewhere.

\section*{Acknowledgements}
The authors would like to thank Prof.~Matej 
Pav\v{s}i\v{c} for bringing to our attention the 
very interesting references~\cite{pavsic1},~\cite{pavsic2},~\cite{pavsic3}.   JBM acknowledges support from CONACYT-Mexico. AM acknowledges financial support from CONACYT-Mexico under project CB-2014-243433.  JMM acknowledges Prof.~Ugo Moschella for discussion at the Escola Patricio Letelier de 
F\'isica-Matem\'atica 2016, Ubu, Brazil, and for bringing references~\cite{moschella1} and~\cite{moschella2} 
  to our attention.

\section*{References}


\end{document}